 \documentclass[final,5p,times,twocolumn]{elsarticle}

 \usepackage{graphics}
 \usepackage{epsfig}

\usepackage{amssymb}

\begin{document}




\title {\bf The Standard Model of Particle Physics. Neutrino Oscillations.}

\author{Giorgio Giacomelli\\ 
Physics Department of the University of Bologna and INFN Sez. of Bologna\\
\vspace{0.3cm}
{\bf Special lecture given at the 24$^{th}$ ICNTS Conference, 5 september 2008, Bologna, Italy.}}

\begin{frontmatter}

\begin{abstract}

 The Standard Model (SM) of Particle Physics was tested to great precision by experiments at the highest energy colliders (LEP, Hera, Tevatron, Slac). The only missing particle is the Higgs boson, which will be the first particle to be searched for at the new Large Hadron Collider (LHC) at CERN. The SM anticipated that there are 3 types of left handed neutrinos. Experiments on atmospheric and solar neutrinos (made in Japan, Italy, Canada, Russia and the US) have shown the existence of neutrino oscillations, which imply that neutrinos have very small mass differences and violate the conservation of individual leptonic numbers. Neutrino oscillations were verified in long baseline neutrino experiments (in Japan and in the USA); and cosmology has given reasonably precise indications on the sum of the neutrino masses. In this paper will be summarized some of the main properties of the SM and some of the main results obtained in the field and the experiments in preparation. Some of the main open questions will be briefly discussed.
\end{abstract}




%

\end{frontmatter}

\section{Introduction. The Standard Model}
\label{intro}

Fig. 1 illustrates the basic components of the Standard Model of Particle Physics [1, 2]: quarks and leptons are the basic components of matter: they are fermions with spin 1$/$2 and may be classified in 3 families: the quarks u, d and the leptons $\nu_{e}$, e$^{-}$ belong to the first family; c, s, $\nu_{\mu}$, $\mu^{-}$ to the second family; t, b, $\nu_{\tau}$, $\tau^{-}$ to the third family (one can say that they have different $flavours$). Only e$^{-}$, d, u are part of ordinary matter; all others are unstable and are produced in high-energy collisions. Fig. 2 shows the mass values of all basic fermions: note that they cover a mass range of $\sim$13 orders of magnitude. In the original formulation of the SM the neutrinos were massless and left handed. We now know that they have very small masses and that they are subject to neutrino oscillations.

Other fundamental objects of the SM are the force carriers, which are bosons of spin 1. They are; the massless photon, responsible of the Electromagnetic Interaction (QED), the 8 massless gluons for the Strong Interaction (QCD) and the heavy Z$^{0}$, W$^{+}$, W$^{-}$, the carriers of the Weak Interaction. 

The total number of basic constituents and carriers is large: 6 quarks, which according to QCD, come in three $colours$ (18), 6 leptons, 12 force carriers (in total 35 particles and one has to consider also the corresponding antiquarks and antileptons (18)). 

The formal theory of the SM is based on the gauge symmetry, which requires zero masses. In order to explain the observed masses one must introduce at least one scalar Higgs boson, which is needed for the spontaneous breaking of the symmetry and the generation of masses. The coupling of the Higgs boson is predicted by the SM, but not its mass [1- 4]. One can only say that its mass should presumable be larger than 130 GeV and smaller than 1TeV. Thus the Higgs will be the first particle to be searched for at the new Large Hadron Collider at CERN, which will produce high-energy $pp$ collisions at c.m. energies up to 14 TeV.

In the following will be discussed the present status of the SM, future studies at LHC, indications of physics beyond the SM, neutrino oscillations in the $\Delta$m$^{2}$ region indicated by the atmospheric neutrino experiments, long baseline neutrino experiments.

\begin{figure}[h]
\begin{center}
{\centering\resizebox*{!}{6cm}{\includegraphics{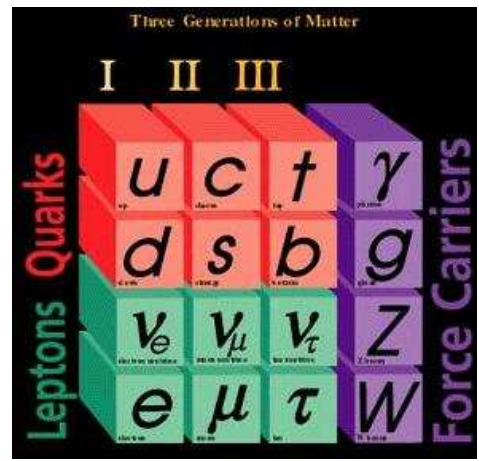}}\par}
\caption{\small The basic elements, leptons and quarks, and the force carriers of the SM.}
\label{fig:1}
\end{center}
\end{figure}

\begin{figure}[h]
\begin{center}
{\centering\resizebox*{!}{6.6cm}{\includegraphics{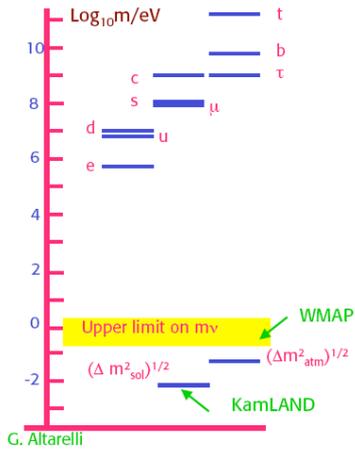}}\par}
\caption{\small Masses of leptons and quarks.}
\label{fig:2}
\vspace{-0.3cm}
\end{center}
\end{figure}

\section{Electroweak fits. QCD}

The SM of particle physics was checked to an unprecedent level of accuracy by the precision measurements made at LEP, SLC, HERA and the Fermilab colliders: Fig. 3 gives the present status of the precision measurements.

The SM has some theoretical inconsistencies and too many parameters; it could be a low energy approximation of a more complete theory. Thus many physicists are looking for physics beyond the SM, for instance supersymmetry, compositeness, etc. New particle searches thus remain an important subject of research. Connected with these problems there is also a strong interest in finding the deep structure of the proton at ever smaller distances.

Neutrino oscillations give indications on the presence of physics beyond the SM, even if part of the properties of neutrino oscillations may be included in the SM. 

The Z$^{0}$ decays predominantly into $q\overline{q}$ pairs, which yields a clean sample of events to test QCD, the theory of the Strong Interaction. The $q\overline{q}$ pair is not observed directly, but it gives rise to two opposite jets of hadrons. Before ``fragmentation'' one of the quarks may radiate a gluon by a process similar to bremsstrahlung yielding 3 jets of hadrons. The ratio of the number of 3-jets to the number of 2-jet events is one way of measuring $\alpha_{s}$, the $strong$ $coupling$ $constant$. This is a foundamental parameter, which was precisely measured with a variety of experiments. They established the $flavour$ $independence$ of $\alpha_{s}$, the running of $\alpha_{s}$, that is its decrease with increassing energy: $\alpha_{s}$(m$_{Z}$)=0.1176$\pm$0.009 (also this is now a precision measurement [3, 4]); also the electromagnetic coupling is not constant; it increases from zero energy ($\alpha_{EM}$ $\sim$1$/$137) to LEP energies ($\alpha_{EM}$ $\sim$1$/$128) [5, 6].

A large variety of phenomenological studies were made on QCD, including the complexity of the hadron spectrum, confinement, phase transitions, etc. [3, 7, 8]. In the past the reference relativistic quantum field theory was QED, but now many physicists consider QCD a better defined theory than QED.

\begin{figure}[h]
\begin{center}
{\centering\resizebox*{!}{8.5cm}{\includegraphics{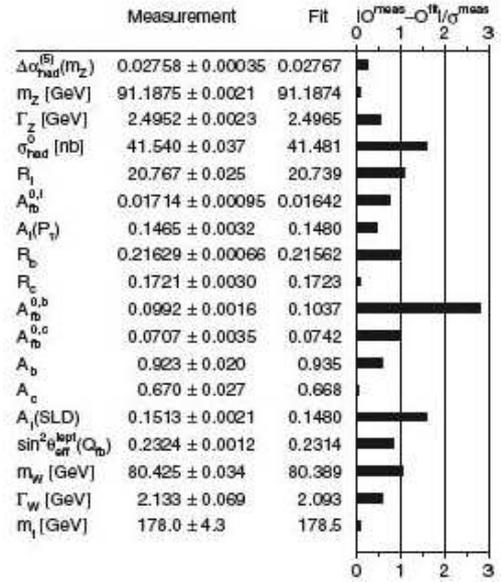}}\par}
\caption{\small Comparison of the results of precision measurements with the expectations of the SM [5].}
\label{tab}
\end{center}
\end{figure}

\section{LHC. Experiments at LHC.}

Fig. 4 shows the CERN complex of accelerators: an Electrostatic proton accelerator, a linear accelerator, a booster proton synchrotron, the 28 GeV Proton Synchrotron (PS), the 400 GeV Super Proton Synchroton (SPS) and finally the Large Hadron Collider (LHC) which should yield proton-proton collisions up to center of mass (cm) energies of 14 TeV. The first accelerators are relative old (and need maintenance) while LHC is a brand new accelerator with helium cooled superconducting magnets.

Fig. 5 shows the SPS supercycle used in september 2008: the first cycle accelerates protons up to 250 GeV which produce secondary beams in the SPS experimental areas; there are then three cycles of protons at 400 GeV, fast extracted to yield neutrinos for the CERN to Gran Sasso beam (CNGS), the last cycle yields 400 GeV protons to the two colliding proton beams of LHC.

The LHC ring is located in an underground tunnel at a depth of $\sim$100 m [9]. In different colliding points are placed the two general purpose detectors ATLAS and CMS (see Fig. 6), then the ALICE detector, designed to study the quark gluon plasma, LHCb for $b$-quark physics, LHCf for forward physics and TOTEM for the measurement of the total $pp$ cross sections.

It may be worth recalling that the main LHC experiments use 4$\pi$ general purpose detectors, with many subdetectors, most of which immersed in a strong magnetic field. They have several hundred thousand electronic channels and very many microprocessors and computers. The detectors have a cylindrical symmetry, with a ``barrel'' and ``endcaps'' structure; some are further structured in different subdetectors. Starting with the innermost detectors and proceeding outward one finds: a microvertex detector, a central tracking detector with dE$/$dx and time-of-flight capabilities; the momentum of produced charge particles is measured by track curvature in the magnetic field. Then follow the electromagnetic and hadron calorimeters, and, after the iron of the magnetic field return yoke, the muon detector. The collider luminosity is measured at each location using forward detectors and eventually precision ``luminometers'' (see for ex. [10, 11]).

\begin{figure}[t]
\begin{center}
{\centering\resizebox*{!}{7.6cm}{\includegraphics{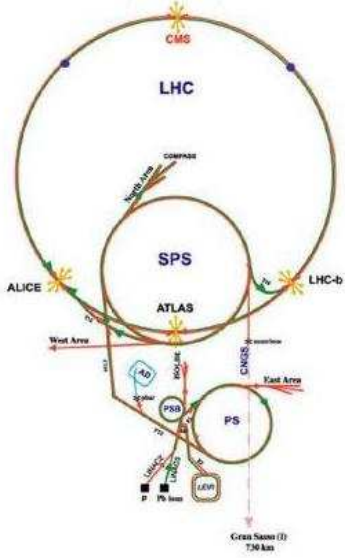}}\par}
\caption{\small Scheme of the CERN complex of Accelerator (not to scale).}
\label{fig:3}
\end{center}
\end{figure}

The experiments at LEP, SLC, Fermilab and HERA, required tens of groups and hundreds of physicists and engineers, with interconnections at the national and regional levels. The experiments at the LHC require each hundreds of groups and thousands of physicists, with interconnections in a sort of world organization {\footnote{ There were discussions on possible catastrophic situations created by the production of mini black holes, nuclearites and$/$or transition to a new vacuum. The conclusions from several analysis teams are that these possibilities are extremely improbable. Moreover the exposure to the highest energy cosmic ray protons during the life of the solar system (equivalent to $\sim$millions of years of LHC runs) did not show any problem. LHC will not destroy our universe and will give us the possibility to know it better.}}.

\section{Neutrinos. Neutrino Oscillations}

If at least two $\nu$'s have non-zero masses, one has to consider the $weak$ $flavour$ $eigenstates$ $\nu_{e}$, $\nu_{\mu}$, $\nu_{\tau}$ and the $mass$ $eigenstates$ $\nu_{1}$, $\nu_{2}$, $\nu_{3}$. Flavour eigenstates are relevant in decays ($\pi^{+}\to\mu^{+}+\nu_{\mu}$) and interactions ($\nu_{\mu}+n\to\mu^{-}+p$), while mass eigenstates are relevant in neutrino propagation. Flavour eigenstates may be written as linear combinations of mass eigenstates. For 2 flavours ($\nu_{\mu}$, $\nu_{\tau}$) and 2 mass eigenstates ($\nu_{2}$, $\nu_{3}$) one writes 
\begin{eqnarray}
\nu_{\mu}=\nu_{2} cos\theta_{23}+\nu_{3} sin\theta_{23}\nonumber \\
\nu_{\tau}=-\nu_{2} sin\theta_{23}+\nu_{3} cos\theta_{23}
\end{eqnarray}
where $\theta_{23}$ is the mixing angle. The survival probability of a $\nu_{\mu}$ ``beam'' is
\begin{equation}
\hspace{-0.8cm} P(\nu_{\mu}\to\nu_{\mu})=1-P(\nu_{\mu}\to\nu_{\tau})=1-sin^{2}2\theta_{23}sin^{2}\left(\frac{1.27\Delta^{2}_{23}L}{E_{\nu}}\right)
\end{equation}
where $\Delta$m$^{2}_{23}$=m$^{2}_{3}$-m$^{2}_{2}$ and L is the distance from $\nu_{\mu}$ production to $\nu_{\mu}$ detection. The simple formula Eq. 2 is modified by additional flavours and by matter effects. Here we shall consider only higher energy $\nu$'s.

\begin{figure}[h]
\begin{center}
{\centering\resizebox*{!}{5cm}{\includegraphics{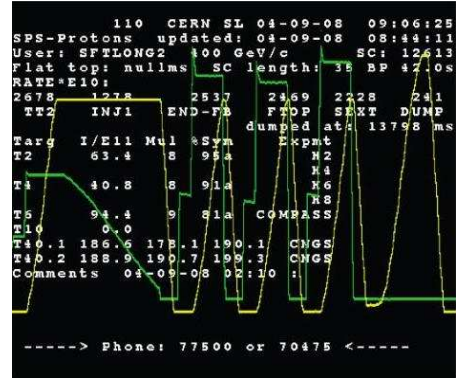}}\par}
\caption{\small SPS Page 1 shows the SPS supercycle: the first cycle is for a test beam at the SPS, 3 cycles for the CNGS beam and one cycle for the LHC.}
\label{fig:4}
\end{center}
\end{figure}

\begin{figure}[h]
\begin{center}
{\centering\resizebox*{!}{5.4cm}{\includegraphics{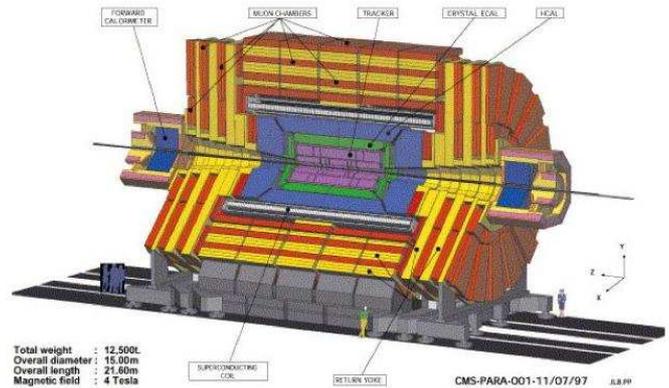}}\par}
\caption{\small CMS. A compact Solenoidal Detector for the LHC.}
\label{fig:5}
\vspace{-0.5cm}
\end{center}
\end{figure}

\section{Atmospheric Neutrinos}

A high-energy primary cosmic ray (CR), proton or nucleus, interacts in the upper atmosphere producing a large number of pions and kaons, which decay yielding muons and $\nu_{\mu}$'s; also the muons decay yielding $\nu_{\mu}$ and $\nu_{e}$. The ratio of the numbers of $\nu_{\mu}$ to $\nu_{e}$ is $\simeq$2 and N$_{\nu}$/N$_{\overline{\nu}}$$\simeq$1. These ``Atmospheric neutrinos are produced'' at 10-20 km above ground, and proceed towards the earth.

\begin{figure}[h]
\begin{center}
\hspace{-0.6cm}
{\centering\resizebox*{!}{4.4cm}{\includegraphics{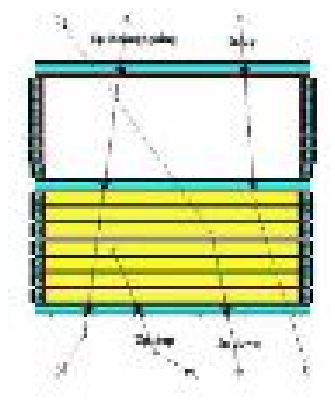}}}
 {\centering\resizebox*{!}{4.4cm}{\includegraphics{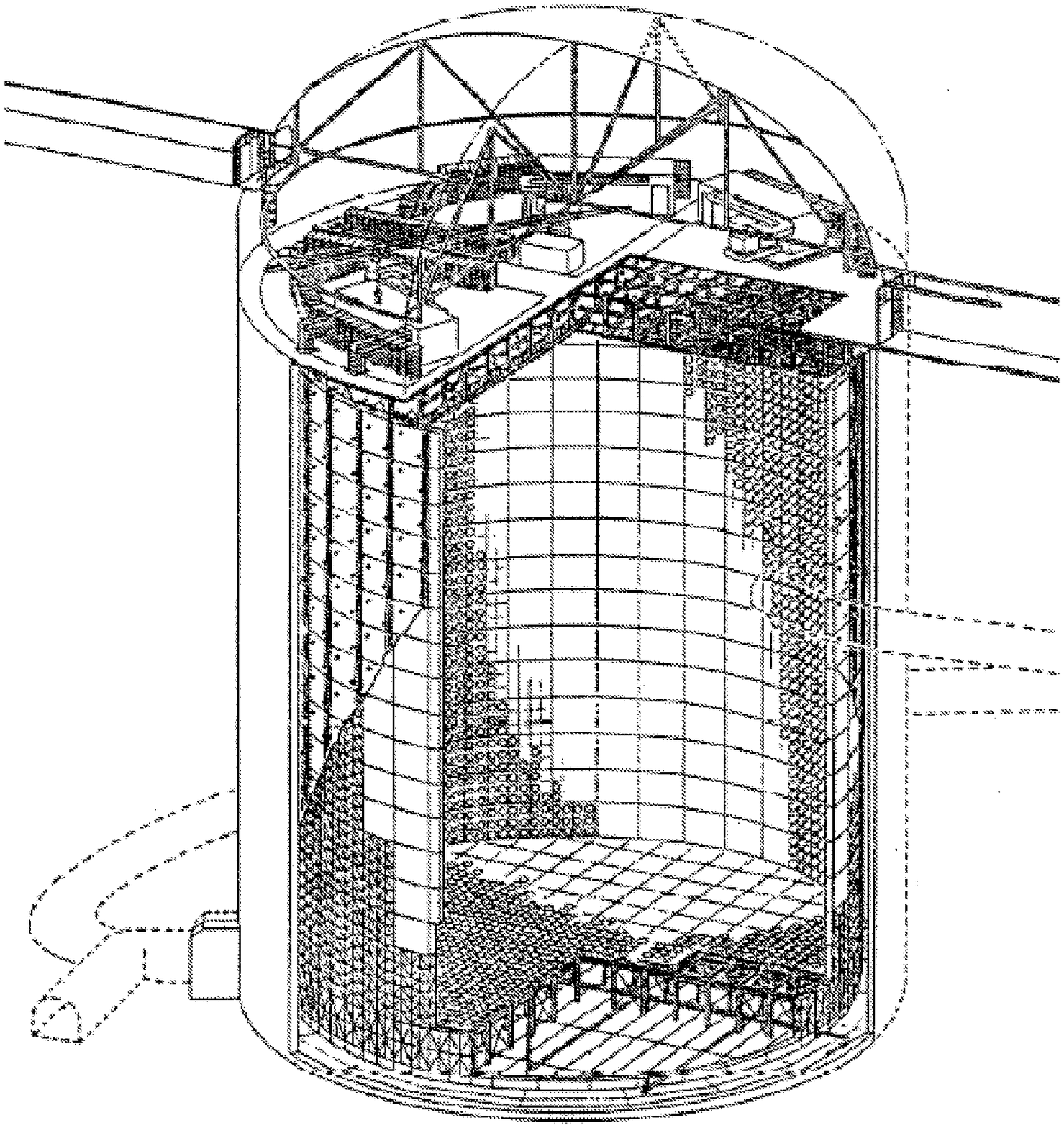}}\par}
\caption{\small (a) Cross-section of the MACRO detector and sketch of event topologies. (b) Schematic layout of the SK detector.}
\label{fig:6}
\vspace{-0.1cm}
\end{center}
\end{figure}

\begin{figure}[h]
\begin{center}
{\centering\resizebox*{!}{6cm}{\includegraphics{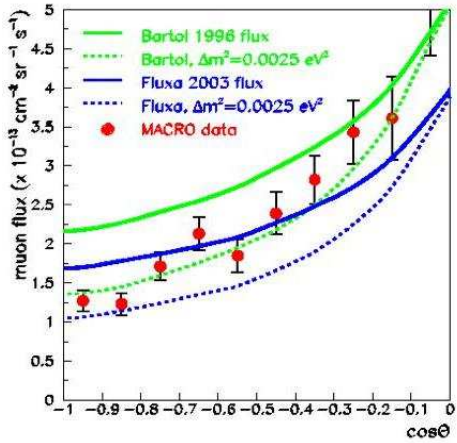}}\par}
\caption{\small MACRO upthroughgoing muons compared with oscillated and non oscillated MC predictions.}
\label{fig:7}
\vspace{-0.6cm}
\end{center}
\end{figure}

 Atmospheric neutrinos are well suited to study neutrino oscillations, since they have energies from a fraction of GeV to more than 100 GeV and travel distances L from few tens of km (downgoing neutrinos) up to 13000 km (upgoing neutrinos); thus L/E$_{\nu}$ ranges from $\sim$1 km/GeV to $\sim$10$^{5}$ km/GeV. With these $\nu$'s one may study $\nu$ oscillations for 10$^{-3}$$<$$\Delta$m$^{2}$$<$10$^{-1}$ eV$^{2}$.

The early water Cherenkov detectors IMB [12] and Kamiokande [13] reported anomalies in the ratio of muon to electron neutrinos, while tracking calorimeters and the Baksan [14] scintillator detector did not find any. In 1995 MACRO found a deficit for upthrougoing muons [15]. Then Soudan 2 [16] confirmed the ratio anomaly. In 1998 Soudan 2 [17], MACRO [18, 19] and SuperKamiokande [20] reported deficits in the $\nu_{\mu}$ fluxes and angular distribution distortions with respect to non oscillated Monte Carlo (MC) predictions; instead the $\nu_{e}$ distributions were in agreement with non oscillated MCs. These features may be explained in terms of $\nu_{\mu}\leftrightarrow\nu_{\tau}$ oscillations.

The atmospheric neutrino flux was computed by many authors in the mid 1990s [21] and in the early 2000s [22]. The last calculations had many improvements, but also a new scale uncertainty, Figs. 8, 9.

{\bf Soudan 2} used a modular fine grained tracking and showering calorimeter of 963 t located in the Soudan Gold mine in Minnesota. The double ratio integrated over $\theta$ is R$'$=(N$_{\mu}$/N$_{e}$)$_{DATA}$/(N$_{\mu}$/N$_{e}$)$_{MC}$=0.68$\pm$0.11$_{stat}$.

\begin{figure}[h]
\begin{center}
{\centering\resizebox*{!}{5.2cm}{\includegraphics{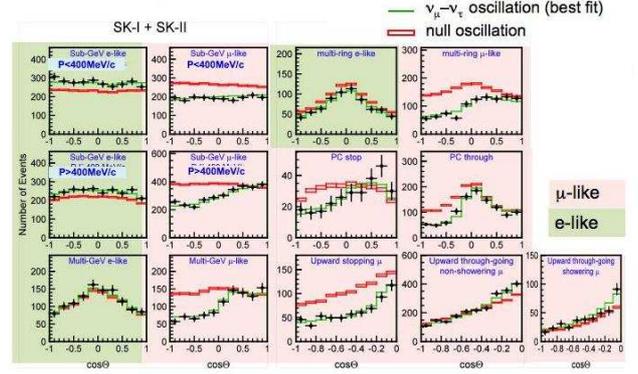}}\par}
\caption{\small Zenith angle distributions from the SK detector (SK-I+SK-II).}
\label{fig:5}
\vspace{-0.5cm}
\end{center}
\end{figure}

{\bf MACRO} (12m$\times$77m$\times$9.5m) at Gran Sasso (GS) detected upgoing $\nu_{\mu}$'s via CC interactions $\nu_{\mu}\leftrightarrow\mu$ using streamer tubes and scintillators. Events were classified as shown in Fig. 7a. A comparison of MACRO data with different MCs is shown in Fig. 8. In order to reduce the effects of systematic uncertainties in the MCs, MACRO eventually used the following 3 independent ratios (it was checked that all MCs yield the same predictions for the ratios):

(i) High-Energy Data:zenith ratio: R$_{1}$=N$_{vert}$/N$_{hor}$

(ii) High and Low En. Data: $\mu$ energy ratio: R$_{2}$=N$_{low}$/N$_{high}$

(iii) Low Energy Data:R$_{3}$=$(Data/MC)_{IU}$/$(Data/MC)_{ID+UGS}$.
The no oscillation hypothesis was ruled out by $\sim$5$\sigma$. Using the 3 ratios, one obtains sin$^{2}$2$\Theta$=1, $\Delta$m$^{2}_{23}$=2.3$\cdot$10$^{-3}$ eV$^{2}$. Using Bartol96, one adds the flux information: 

(iv) High en. $\mu$ (scale error $\simeq$17$\%$): R$_{4}$=N$_{meas}$/N$_{MC}$.

(v) Low en. muons (scale error $\sim$21$\%$): R$_{5}$=N$_{meas}$/N$_{MC}$.
These two ratios leave the best fit values unchanged and improve the significance to $\sim$6$\sigma$.

{\bf SuperKamiokande (SK)} is a large cylindrical water Cherenkov detector containing 50 kt of water (fiducial mass 22.5 kt); the light is seen by 50-cm-diameter inner-facing phototubes (PMTs), Fig. 7b. The 2m thick outer layer of water acts as an anticoincidence using smaller outward facing PMTs. The large detector mass allows to collect high statistics of $fully$ $contained$ events (FC), divided into $sub-GeV$ and $multi-GeV$ events, with energies below and above 1.33 GeV. $Multi-ring$ events are treated as a separate category. The $partially$ $contained$ events are CC interactions with vertex within the fiducial volume and at least a charged particle, the $\mu$, exits the detector. $Upward-going$ $muons$, produced by $\nu_{\mu}$ from below interacting in the rock, are divided into $stopping$ ($\langle E_{\nu}\rangle \sim7 GeV$) and throughgoing ($\langle E_{\nu}\rangle \sim70\div80 GeV$). The zenith distributions for $e$-like and $\mu$-like $sub-GeV$ and $multi-GeV$ events are shown in Fig. 9 left and right, respectively.

The MC problem exists also in SK: the $e$-like events were in agreement with the 1995 MC predictions for no-oscillations. For $e$-like events, the new MC predictions are low: to reduce these problems the normalization is left as a free parameter. The overall best fit yields for $\nu_{\mu}\rightarrow\mu$ maximal mixing and $\Delta$m$^{2}_{23}$=2.5$\cdot$10$^{-3}$eV$^{2}$.

{\bf Exotic oscillations.} MACRO and SK data were used to search for sub-dominant oscillations due to a possible Lorentz invariance violation (there would be mixing between flavour and velocity eigenstates). Limits were placed in the Lorentz violation parameter $\vert$ $\Delta$ v $\vert$ $<$ 6 $\cdot$ $10^{-24}$ at $sin^{2}2\theta_{\nu}$=0 and $\vert$ $\Delta$ v $\vert$ $<$ 4 $\cdot$ $10^{-26}$ at $sin^{2}2\theta_{\nu}$=$\pm 1$ [23]. 

{\bf Neutrino decay} could be an explanation for $\nu$ disappearance; no radiative decay was observed [24, 25].
 
\section{Long Baseline $\nu$ Experiments}

Long baseline $\nu$ experiments allow further insight into $\nu$ physics. The first long baseline $\nu$ beam was the KEK to Kamioka (K2K) beam, the 2$^{nd}$ was the Fermilab to the Soudan mine beam (NuMi). {\bf MINOS}, on the NuMi low energy $\nu$ beam, is a large magnetised steel scintillator tracking calorimeter, complemented by a similar near detector. It confirmed the atmospheric $\nu$ oscillation picture with maximal mixing and $\Delta$ m$^{2}_{23}$=2.38$\cdot$10$^{-3}$ eV$^{2}$. 
 
The CERN to Gran Sasso, CNGS [26], was tried for short periods in 2006 and  2007. The main components of the $\nu_{\mu}$ beam at CERN are the 400 GeV proton beam from the SPS transported to an underground target. Secondary pions and kaons are focused into a parallel beam by 2 magnetic lenses, called $horn$ and $reflector$. Pions and kaons decay into $\nu_{\mu}$ and $\mu$ in a long decay pipe. The remaining hadrons are absorbed in the hadron stop. The $\mu$'s are monitored in 2 muon detectors.

Fig. 10 shows the path of the CNGS $\nu_{\mu}$s from CERN to GS. It also shows the synchronization via GPS of the atomic clocks at CERN and GS. Fig. 5 shows the scheme of the SPS operation in 2008. The $\nu$ beam is optimised for producing a maximum number of CC $\nu_{\tau}$ interactions in OPERA. The mean $\nu_{\mu}$ energy is $\sim$17 GeV, the $\nu_{\mu}$ contamination $\sim$2$\%$, background is $<$1$\%$. The muon beam size at the 2$^{nd}$ muon detector at CERN is $\sigma\sim$1m; the $\nu_{\mu}$ beam size at GS is $\sigma\sim$1 km. The first low intensity test beam was sent to GS in August 2006 and 3 detectors (OPERA, LVD and Borexino) obtained their first events {\footnote {The ICARUS experiment is setting up and should run in 2010}}. The low intensity CNGS was stable and of high quality. The SPS sent a pulse of 2 neutrino bursts, each of 10.5 $\mu$s duration, separated by 50 ms, every 12 s. A higher intensity beam did not happen because of a water leak at CERN. In 2007 a 2$^{nd}$ test was successful, but the high intensity run was cancelled because of cooling problems. In 2008 the beam run properly (see below).

At GS, the CNGS beam is seen by:
 
{\bf Borexino}, in Hall C, is an electronic detector designed to study solar $\nu_{e}$'s coming from Be$^{7}$ decays in the sun [27]. 

{\bf LVD}, an array of liquid scintillators with a mass of 1000 t, designed to search for $\overline{\nu_{e}}$'s from gravitational stellar collapses [28]. LVD, in Hall A, is a neutrino flux monitor.

{\bf OPERA} [29], in Hall C, is a hybrid emulsion-electronic detector, designed to search for $\nu_{\mu}\leftrightarrow\nu_{\tau}$ oscillations in appearance mode in the parameter region indicated by the atmospheric neutrinos, K2K and MINOS. The $\nu_{\tau}$ appearance will be made by direct detection of the $\tau$ lepton, from $\nu_{\tau}$ CC interactions and the $\tau$ lepton decay products. To observe the decays, a spatial resolution of $\sim1 \mu$m is necessary; this is obtained in emulsion sheets interspersed with thin lead target plates (Emulsion Cloud Chamber (ECC)). OPERA may also search for the subleading $\nu_{\mu}\leftrightarrow\nu_{e}$ oscillations and make a variety of other observations with its electronic detectors. The detector, Figs. 11, 12, is made of two identical supermodules, each consisting of a $target$ $section$ with 31 target planes of lead/emulsion modules (``bricks''), of $a$ $scintillator$ $tracker$ $detector$ and a muon spectrometer. An $anticoincidence$ $wall$ separates muons coming from interactions in OPERA from those in the rock.

The $muon$ $spectrometer$ consists of 2 iron magnets instrumented with $Resistive$ $Plate$ $Chambers$ (RPC) and $drift$ $tubes$. Each magnet is an 8$\times$8 m$^{2}$ dipole with a field of 1.52 T in the upward direction on one side and in the downward direction on the other side. A $precision$ $tracker$ measures the muon track coordinates in the horizontal plane with drift tubes, placed in front and behind each magnet and between the 2 magnets. The muon spectrometer has a $\Delta p/p \leq$0.25 for muon momenta $\leq$25 GeV$/$c. Two 45$^{\circ}$ crossed planes of $glass$ $RPC's$ ($XPC's$) are installed in front of the magnets.

The basic target module is a ``$brick$'', consisting of 56 lead plates (1 mm thick) and 57 emulsion layers. A brick has a size of 10.2$\times$12.7 cm$^{2}$, a depth of 7.5 cm (10 radiation lengths) and a weight of 8.3 kg. Two additional emulsion sheets, the $changeable$ $sheets$ (CS), are glued on its downstream face. The bricks are arranged in walls. Within a brick, the achieved spatial resolution is $<$1 $\mu$m and the angular resolution is $\sim$2 mrad. Walls of target trackers provide the $\nu$ interaction trigger and identify the brick in which the interaction took place.

The bricks were made by the $Brick$ $Assembly$ $Machine$ (BAM) and are handled by the $Brick$ $Manipulator$ $System$ (BMS).

A fast automated scanning system with a scanning speed of $\sim$20 cm$^{2}/$h per emulsion (each 44 $\mu$m thick) is needed to cope with daily analyses of many emulsions. This is a factor of $>$10 increase with respect to past systems. For this purpose were developed the $European$ $Scanning$ $System$ (ESS) [30] and the Japanese $S-UTS$. An emulsion is placed on a holder and the flatness is guaranteed by a vacuum system. By adjusting the focal plane of the objective, 16 tomographic images of each field of view are taken at equally spaced depths. The images are digitized, sent to a vision processor, analyzed to recognize sequences of aligned grains. The 3-dimensional structure of a track in an emulsion layer ($microtrack$) is reconstructed by combining clusters belonging to images at different levels. Each microtrack pair is connected across a plastic base to form the $base$ $track$. A set of base tracks forms a $volume$ $track$.

In the 2006 test run OPERA made a study of the $\Theta$ angle distribution of events on-time with the beam; this yielded a mean $\mu$ vertical angle of 3.4$^{\circ}$ in agreement with expectations for $\nu_{\mu}$ originating from CERN and travelling under the earth surface to the GS underground halls. A second test proved the capability of going from the centimetre scale of the electronic tracker to the micrometric resolution of nuclear emulsions [29]. 

During the short 2007 test run, interactions in the lead bricks were seen by the electronic detectors, confirmed by tracks in the CS; the event vertexes were observed in the emulsions analysed by microscopes: Fig. 12a shows an interaction as seen in the electronic detectors, Fig. 12b the vertex region in the emulsions. The test confirmed the validity of the methods to associate electronic detectors to nuclear emulsions.

The 2008 run used the full detector and a reasonable CNGS total intensity ($\sim$1.9$\cdot$10$^{18}$ pot).
 
\begin{figure}[h]
\begin{center}
{\centering\resizebox*{!}{4.95cm}{\includegraphics{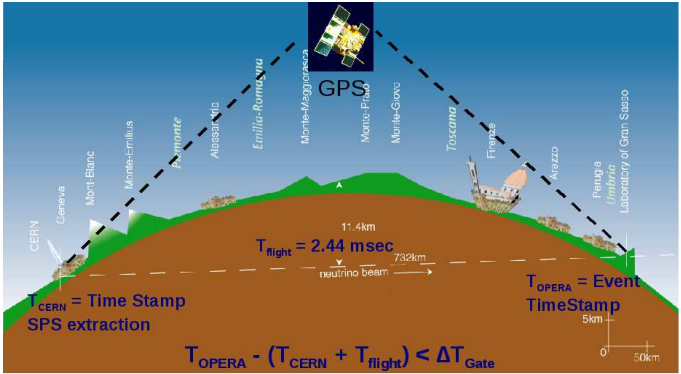}}\par}
\caption{\small Sketch of the 730 km neutrino path from CERN to Gran Sasso and the GPS selection of events.}
\label{fig:9}
\end{center}
\end{figure}

\begin{figure}[h]
\begin{center}
{\centering\resizebox*{!}{4.5cm}{\includegraphics{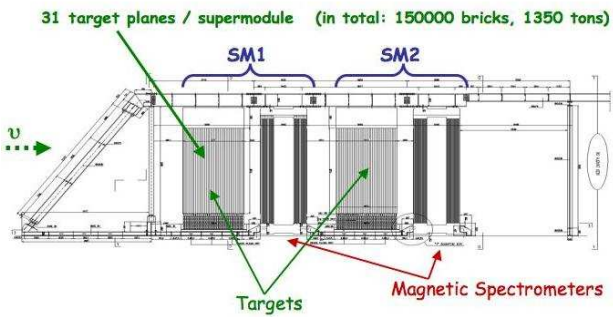}}\par}
\caption{\small Schematic side view of the OPERA experiment.}
\label{fig:10}
\end{center}
\end{figure}

\section{Conclusions. Outlook}

The Standard Model of Particles Physics needs further confirmation: the Higgs boson is probably the main open problem of particle physics. The Higgs will certainly be searched for at the LHC using a variety of methods. The search for new physics beyond the SM will be another main field of research at LHC; Supersymmetry remains one of the theoretically favourite topics and this includes dark matter. But one should continue the searches for Compositeness, Technicolor, Extra dimensions and for other topics already searched for at lower energy colliders. 

\begin{figure}[h]
\begin{center}
{\centering\resizebox*{!}{3.8cm}{\includegraphics{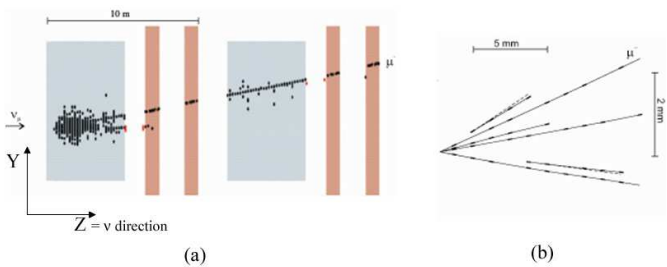}}\par}
\caption{\small (a) Side view of a CC event as seen in the OPERA electronic detectors. (b) The event vertex as reconstructed in the emulsions.}
\label{fig:11}
\end{center}
\end{figure}

In the field of Astroparticle Physics one should continue the study of neutrino physics and astrophysics. In particular: is the $\nu$ a Majorana fermion? and continue Dark Matter searches.

The atmospheric neutrino anomaly became, in 1998, the atmospheric neutrino oscillation hypothesis with maximal mixing and $\Delta m^{2}_{23}$$\simeq$2.4$\cdot$10$^{-3}$ eV$^{2}$. It was later confirmed with more data and by the first two long baseline experiments. All experiments agree on maximal mixing, while the $\Delta m^{2}_{23}$ values are: Soudan-2 5.2$\cdot$10$^{-3}$ eV$^{2}$, MACRO 2.3 , SK 2.5 , K2K 2.7 , MINOS 2.38$\cdot$10$^{-3}$ eV$^{2}$. These results come from disappearance experiments ($\nu_{\mu}\leftrightarrow\nu_{\mu}$). An appearance experiment ($\nu_{\mu}\leftrightarrow\nu_{\tau}$) with $\nu_{\tau}$ detection may solve conclusively the situation. [31].

{\bf Acknowledgements.} We acknowledge discussions with many colleagues; we thank Drs. M. Errico, M. Giorgini and V. Togo for their cooperation.

\end{document}